\documentclass[usenatbib]{mn2e}
\usepackage{natbib}
\usepackage{graphicx}

\title[Spectroscopic survey of the Galaxy with {\it Gaia}]{Spectroscopic
survey of the Galaxy with {\it Gaia}\\ I. Design and performance of the
Radial Velocity Spectrometer}
\author[D. Katz, U.~Munari, M.~Cropper et al.]{D. Katz$^{1}$\thanks{E-mail: david.katz@obspm.fr},
U.~Munari$^{2}$,       M.~Cropper$^{3}$,
T.~Zwitter$^{4}$,      F.~Th\'evenin$^{5}$,     M.~David$^{6}$,
\newauthor
Y.~Viala$^{1}$,        F.~Crifo$^{1}$,          A.~Gomboc$^{7, 4}$,
F.~Royer$^{8, 1}$,     F.~Arenou$^{1}$,         P.~Marrese$^{2}$,
\newauthor
R.~Sordo$^{2}$,        M.~Wilkinson$^{9}$,      A.~Vallenari$^{10}$,
C.~Turon$^{1}$,        A.~Helmi$^{11}$,         G.~Bono$^{12}$,
\newauthor
M.~Perryman$^{13}$,    A.~G\'omez$^{1}$,        L.~Tomasella$^{2}$, 
F.~Boschi$^{2}$,       D.~Morin$^{1}$,          M.~Haywood$^{1}$,
\newauthor
C.~Soubiran$^{14}$,    F.~Castelli$^{15, 16}$,  A.~Bijaoui$^{5}$,
G.~Bertelli$^{10}$,    A.~Prsa$^{4}$,           S.~Mignot$^{1}$,
\newauthor
A.~Sellier$^{1}$,      M.-O.~Baylac$^{1}$,      Y.~Lebreton$^{1}$,
U.~Jauregi$^{4}$,      A.~Siviero$^{2}$,        R.~Bingham$^{17, 3}$,
\newauthor
F.~Chemla$^{1}$,       J.~Coker$^{3}$,          T.~Dibbens$^{3}$,
B.~Hancock$^{3}$,      A.~Holland$^{18}$,       D.~Horville$^{1}$,
\newauthor
J.-M.~Huet$^{1}$,      P.~Laporte$^{1}$,        T.~Melse$^{1}$,
F.~Say\`ede$^{1}$,     T.-J.~Stevenson$^{18}$,  P.~Vola$^{1}$,
\newauthor
D.~Walton$^{3}$      and B.~Winter$^{3}$.\\
$^{1}$Observatoire de Paris, GEPI, 5 Place Jules Janssen, F-92195
Meudon France\\
$^{2}$Osservatorio Astronomico di Padova INAF, sede di Asiago, 36012 Asiago
(VI), Italy\\
$^{3}$Mullard Space Science Laboratory, University College London, Holmbury St
Mary, Dorking, Surrey RH5 6NT, UK\\
$^{4}$University of Ljubljana, Dept. of Physics, Jadranska 19, 1000 Ljubljana,
Slovenia\\
$^{5}$O.C.A., BP 4229, F-06304 Nice Cedex 4, France\\
$^{6}$University of Antwerp, Middelheimlaan 1, B-2020 Antwerpen, Belgium\\
$^{7}$Astrophysics Research Institute, Liverpool John Moores University, 
Twelve Quays House, Egerton Wharf, Birkenhead, CH41 1LD, UK\\
$^{8}$Observatoire de Gen\`eve, 51 chemin des Maillettes, CH-1290 Sauverny\\
$^{9}$Institute of Astronomy, Madingley Road, cambridge CB3 0HA, UK\\
$^{10}$INAF - Osservatorio Astronomico di Padova, Vicolo Osservatorio 5,
35122 Padova, Italy\\
$^{11}$Kapteyn Astronomical Institute, P.O.Box 800, 9700 AV Groningen,
Netherlands\\
$^{12}$INAF - Rome Astronomical Observatory, Via Frascati 33, 00040 Monte
Porzio Catone, Italy\\
$^{13}$Research and Scientific Support Department of ESA, ESTEC, Postbus 299,
Keplerlaan 1, Noordwijk NL-2200 AG, the Netherlands\\
$^{14}$Observatoire Aquitain des Sciences de l'Univers, UMR 5804, 2 rue de
l'Observatoire, 33270 Floirac, France\\
$^{15}$CNR-Istituto di Astrofisica Spaziale e Fisica Cosmica, Via del Fosso
del Cavaliere, 00133, Roma, Italy\\
$^{16}$INAF-Osservatorio Astronomico di Trieste, via G.B. Tiepolo 11, 34131
Trieste, Italy\\
$^{17}$Optical Design Service, Cambridge, 29 Millington Road,
Cambridge CB3 9HW, UK\\
$^{18}$University of Leicester, Space Research Centre, University Road,
Leicester, UK}
\begin{document}

\date{Accepted . Received ; in original form }

\pagerange{\pageref{firstpage}--\pageref{lastpage}} \pubyear{2004}

\maketitle

\label{firstpage}

\begin{abstract}
The definition and optimisation studies for the {\it Gaia} satellite
spectrograph, the Radial Velocity Spectrometer (RVS), converged in
late 2002 with the adoption of the instrument baseline. This paper
reviews the characteristics of the selected configuration and presents
its expected performance.  The RVS is a 2.0~$\times$~1.6 degree
integral field spectrograph, dispersing the light of all sources
entering its field of view with a resolving power $R = \lambda /
\Delta \lambda = 11 500$ over the wavelength range [848, 874] nm. The
RVS will continuously and repeatedly scan the sky during the 5 years
of the {\it Gaia} mission.  On average, each source will be observed
102 times over this period.  The RVS will collect the spectra of about
100-150 million stars up to magnitude $V \simeq 17-18$. At the end of
the mission, the RVS will provide radial velocities with precisions of
$\sim$2 km~s$^{-1}$ at $V=15$ and $\sim15-20$ km~s$^{-1}$ at $V=17$,
for a solar metallicity G5 dwarf.  The RVS will also provide
rotational velocities, with precisions (at the end of the mission) for
late type stars of $\sigma_{v \sin i} \simeq 5$ km~s$^{-1}$ at $V
\simeq$ 15 as well as atmospheric parameters up to $V \simeq 14-15$.
The individual abundances of elements such as Silicon and Magnesium,
vital for the understanding of Galactic evolution, will be obtained up
to $V \simeq 12-13$.  Finally, the presence of the 862.0 nm Diffuse
Interstellar Band (DIB) in the RVS wavelength range will make it
possible to derive the three dimensional structure of the interstellar
reddening.
\end{abstract}


\section{Introduction}
Understanding the Milky Way and its neighbourhood, both as a template
of the formation and evolution of galaxies and as a laboratory of
stellar physics, is one of the great astrophysics challenges at the
beginning of the 21st century. The {\it Gaia} satellite
\citep{perryman2001} was conceived and designed to constitute a major
leap forward in this field.  {\it Gaia} was selected as a Cornerstone
of the European Space Agency (ESA) Horizon 2000+ programme in October
2000 and confirmed as part of the ESA Cosmic Vision programme in May
2002 and again in November 2003. The satellite launch is scheduled to
take place ``no later than 2012'' and all the current scientific and
industrial studies are consistent with a launch date in mid-2010.

{\it Gaia}'s payload consists of 3 instruments: an astrometric
instrument, a multi-band photometer (observing in 5 broad and 11
medium-band filters) and a spectrometer which will continuously and
repeatedly scan the sky during the 5 years of the mission. On average,
102 spectra will be obtained per source. The astro-photometric
``sample'' will be complete up to the {\it Gaia} magnitude G=20 ($G-V
\simeq$ $-$0.36 for a solar metallicity G5V star) and will contain
more than a billion stars from the Galaxy and the Local Group.  The
huge size of the sample (about 1\% of the entire stellar content of
the Milky Way galaxy) and the accuracy and wealth of information
collected about the spatial, kinematic and chemical distributions of
the stars will allow {\it Gaia} to decipher many aspects of the origin
and evolution of our Galaxy.  Moreover, given the very large number of
stars surveyed, even the briefest stages of stellar evolution have a
very high probability of being observed by {\it Gaia}.  {\it Gaia}
will also detect and characterise several thousand extra-solar
planetary systems. It will observe some $10^5$ to $10^6$ minor bodies
in the solar system, thousands of galaxies, some $5\times 10^5$
quasars, about 10$^5$ extragalactic supernovae and will provide an
accurate determination of several fundamental physical and
cosmological parameters (e.g. $\gamma$, $\dot{G}$/$G$, $\Omega_M$ or
$\Omega_\Lambda$).

The initial concept of {\it Gaia} was developed around the astrometric
instrument. However, based on the experience of {\it Hipparcos}
\citep{gerbaldi1989, mayor1989, binney1997}, it was clear that the
capability of making radial velocity measurements would be central to
the successful attainment of the main {\it Gaia} science goals. By
supplying the third component of the velocity vector (the other two
being provided by the tangential motions), the velocity along the line
of sight is crucial for the our understanding of the kinematics and
dynamics of the Galaxy. Radial velocities are also necessary to
correct the astrometric data for perspective acceleration (an apparent
displacement on the sky induced by line-of-sight motion and which
varies quadratically with time): simulations predict that the proper
motions and positions of $\simeq 10^5$ stars (unknown {\it a priori})
would be significantly biased if this effect was not removed.
Finally, multi-epoch radial velocities are extremely valuable for the
detection and characterisation of multiple systems and stellar
variability. As {\it Gaia} was taking shape, several studies were
investigating the feasibility, merits and drawbacks of the acquisition
of radial velocities from the ground and from space
\citep{per95,fav95}. In 1997, the decision was made to include a
spectrograph on board {\it Gaia}.

A first concept of the spectrograph was designed in 1998 by Matra
Marconi Space (now Astrium-SAS, Toulouse). Following the approval of
{\it Gaia}, ESA established, in June 2001, sixteen scientific working
groups to support the development of the satellite and to prepare for
the scientific analysis of the data. One of those groups was charged
with the optimisation of the spectrograph, named the ``Radial Velocity
Spectrometer'' (RVS), with respect to the {\it Gaia} science case.
During the following 18 months, the RVS Working Group, with the
support of Astrium, reviewed and refined the scientific objectives and
priorities for the spectrograph, assessed its performance and compared
the advantages and disadvantages of several potential RVS
configurations. The scientific requirements and the technical
constraints converged toward a refined scientific case and a new
spectrograph concept in November 2002. The main modification with
respect to the Matra Marconi design is the increase of the resolving
power. A broader scientific case has also been associated with the
revised configuration of the spectrograph.  The acquisition of
sufficiently accurate radial velocities for the largest possible
sample of stars remains the key scientific objective of the RVS.
Radial velocities are fundamental for the proper understanding of the
Milky Way structure, origin and history. However, the increased
resolution will make it possible, without affecting the radial
velocity performance, to address a wide variety of issues in stellar
physics: binarity, evolution, rotation, pulsation and variability,
atmospheric chemistry, mixing processes and abundance peculiarities.

The RVS science case and general characteristics (e.g. spectral
resolution, wavelength range) have now been defined.  The more
detailed design features (such as the number and type of CCDs) remain
to be confirmed.  This paper presents the revised and optimised
baseline of the Radial Velocity Spectrometer: instrument concept
(\S~\ref{concept}), optical (\S~\ref{optics}) and focal plane
(\ref{focal}) characteristics. Although the instrument concept is
expected to evolve in future phases, it should remain broadly within
the parameters described here. This paper also presents the RVS
calibration (\S~\ref{calibration}) and observational
(\S~\ref{observation}) strategies. Finally, it reviews the stellar and
interstellar parameters that will be provided by the RVS spectra,
namely, radial (\S~\ref{radial}) and rotational (\S~\ref{rotational})
velocities, atmospheric parameters and individual abundances
(\S~\ref{atmospheric}) as well as interstellar reddening
(\S~\ref{interstellar}).

The {\it Gaia} spectroscopic survey is expected to lead to major
improvements in our understanding of the kinematical, dynamical and
chemical structure and history of the Milky Way, as well as in our
comprehension of stellar physics and evolution. Moreover, it will be
ideally suited to the identification and characterisation of multiple
systems. The expected scientific harvest is discussed in a companion
paper \citep{wvt04}.

\section{RVS design}
\subsection{Instrument concept} \label{concept}
The Radial Velocity Spectrometer is an integral field spectrograph: it
uses neither slits, nor fibres, but disperses all of the light
entering its $2.00^\circ\times1.61^\circ$ field of view. As with the
other {\it Gaia} instruments, it will continuously and repeatedly scan
the sky, observing each object, on average, at 102 successive epochs
(see \S~\ref{observation}).

The choice of the spectrograph wavelength range ([848, 874] nm:
\citeauthor{munari1999}~\citeyear{munari1999}) was motivated by a
number of considerations. It is close to the peaks of the energy
distributions of the RVS principal targets: G and K type stars.  In F,
G and K stars, there are three strong core-saturated ionised Calcium
lines in this range, which allows the measurement of radial velocities
even at very low signal to noise ratios ({\it i.e.}  $\sigma Vr
\simeq$~15~km~s$^{-1}$ at $S/N \simeq 1$ per pixel, for a K1V type
star - see \S~\ref{radial}) as well as in very metal poor stars. In
early type stars, this spectral region is dominated by lines of the
Hydrogen Paschen series, which are visible even in very rapidly
rotating stars. It contains a Diffuse Interstellar Band (DIB), located
at 862.0~nm, which appears to be a reliable tracer of the interstellar
reddening (\S~\ref{interstellar}): this will be used, together with
the photometric data, to derive a three-dimensional map of Galactic
interstellar extinction.  In addition, the extinction in the RVS
domain will roughly be a factor of two smaller than in the V
band. This will allow the probing of the Galactic disk over greater
distances than would be feasible at 550 nm.  Finally, this spectral
range is almost free of telluric absorption features and is therefore
usable from the ground for pre-launch preparatory (and possible
complementary, post-mission) observations (\S~\ref{spectra}).

In its original conception, the spectrograph had an effective
resolving power $R = \lambda / \Delta \lambda = 5750$. During the 18
months of the optimisation phase (June 2001 - November 2002) the RVS
working group compared the merits and drawbacks of a broad range of
resolutions from $R = 5000$ to 20000.  The obvious advantage of the
``high'' resolutions is that they carry more spectral information than
the ``low'' ones. The ``weak'' lines are sharper, the contrast between
lines and continuum is better and the rate of line blending is lower.
Therefore, the high resolution spectra provide more accurate
diagnostics for the determination of stellar parameters (e.g. radial
and rotational velocities, atmospheric parameters -- see
\S~\ref{stellar}) of most stars except the ``faintest''\footnote{The
total read-out and background noise, summed over the 26 nm of the
spectrum, increases with the number of samples and therefore with the
resolution. For the very faint stars, the advantages from the larger
amount of spectral information is offset by the higher noise levels.}.
Moreover, for $R \geq 10000$, numerous lines contained in the RVS
infra-red wavelength range are unblended and it becomes possible to
determine the individual abundances of several chemical species and in
particular of alpha elements (e.g. Magnesium, Silicon - see
\S~\ref{atmospheric}).  However, the ``high'' resolutions also present
a drawback. The RVS is an integral field spectrograph. As a
consequence, in regions of high stellar density, the spectra of
neighbouring sources will overlap and the mean rate of overlap grows
linearly with resolution. Numerical simulations have shown that it
will be possible, to a certain extent, to deconvolve the stacked
spectra \citep{zwitter2003} -- see \S~\ref{radial}. Even so, in very
crowded areas the faintest sources will be lost in the noise. In those
very dense regions, the limiting magnitude that the RVS can reach is a
decreasing function of the resolution.

After extensive consideration within the working group of the
advantages and disadvantages of resolving powers between $R = 5000$
and 20000, a resolving power $R = 11500$ was shown to be optimal for
the fulfilment of {\it Gaia}'s scientific objectives. This resolution
will allow the radial velocities of late-type stars to be determined
with a precision $\sigma V_r \leq$~10-20 km~s$^{-1}$ up to $V
\simeq$~17-18 (reaching a precision of the order of 1~km~s$^{-1}$ or
better for the brightest targets), over about 90\% of the sky
(Sect~\ref{radial}).  The rate of spectral overlap will be lower at $R
=$~11500 than at higher resolution, thereby allowing the RVS to probe
fields of higher stellar density.  Moreover, with the resolution $R
=$~11500 it will be possible to study the chemical pattern of the
sources up to $V\simeq$ 12-13 (Sect~\ref{atmospheric}), while a lower
resolution $R =$~5750 did not allow the derivation of individual
abundances because of too much line blending and insufficient contrast
between lines and continuum.  Table~\ref{tab:concept} summarises the
global characteristics of the Radial Velocity Spectrometer.

\begin{table}
\caption{Summary of the RVS general characteristics.}
\begin{center}
\begin{tabular}{l c l} \hline
Spectrograph type               & $\ \ $ & Integral field         \\
Observing mode                  &        & Scan mode              \\
Field of view size              &        & $2.00^\circ \times 1.61^\circ$\\
Wavelength range                &        & [848, 874] nm          \\
Resolving power ($R = \lambda / \Delta \lambda$) &        & 11500 \\
Dispersion orientation          &        &  along scan            \\
Integration time (per transit)  &        & 99 s                   \\ \hline
\end{tabular}
\end{center}
\label{tab:concept}
\end{table}

\subsection{Optics} \label{optics}
\noindent
{\it Gaia} will scan the sky with three telescopes. Two of them
illuminate the astrometric focal plane (which also contains the
broad-band photometric detectors). The third one, the Spectro
telescope, is shared by the Radial Velocity Spectrometer and the
medium-band photometer. The viewing directions of the three telescopes
are located in the same plane.

The Spectro telescope which illuminates both the RVS and the
medium-band photometer is made of three Silicon-Carbide (Si-C),
off-axis, rectangular mirrors. It has an entrance pupil of 0.25 square
metres and a focal length of 2.1 m. The 11 medium-bands of the
photometer cover a wavelength range from the near UV ($\simeq$300~nm)
to the near infra-red ($\simeq$1000~nm). The mirrors are coated with
Aluminium which, unlike other metals, exhibits a reflectivity greater
than $85$\% over the whole photometer wavelength range\footnote{The
reflectivity of Silver drops very rapidly below $\simeq$350~nm}.

\begin{figure}
\resizebox{\hsize}{!}
{\rotatebox{0}{\includegraphics{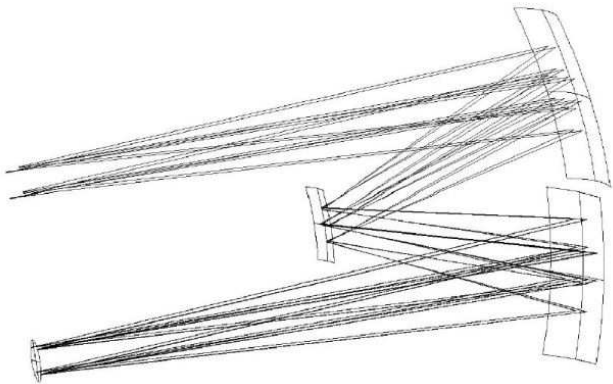}}}
\caption{First optical configuration: Offner relay. The spectrograph
entrance plane (telescope focal plane) is located in the upper left
corner. The RVS focal plane is represented by the ellipse in the lower
left corner. The dispersive element, a grating, is ruled on the
secondary mirror (centre).}
\label{fig:offner}
\end{figure}

\begin{figure}
\resizebox{\hsize}{!}
{\rotatebox{0}{\includegraphics{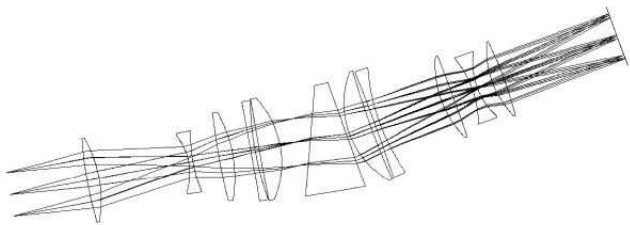}}}
\caption{Second optical configuration: dioptric system. The
spectrograph entrance plane (telescope focal plane) is on the left
side of the figure. The RVS focal plane is represented by the thin
line on the right of the figure.}
\label{fig:dioptric}
\end{figure}

Two optical configurations remain under consideration for the RVS.
The first, proposed by R. Bingham, is based on the Offner Relay
concept (see \citet{reininger1994} for earlier work on this concept).
It is made of three mirrors: two concave (the primary and tertiary
mirrors) and one convex (the secondary mirror).  The wavelengths are
dispersed by a grating ruled on the secondary mirror.  The second
optical configuration being considered follows a classical arrangement
of (i) a dioptric collimator, which parallelises the light rays coming
from the same field point so that they enter the disperser with the
same incident angle, (ii) a dispersive element: a grism; (iii) a
dioptric camera, which refocuses the dispersed image on the
detectors. Fig.~\ref{fig:offner} and Fig.~\ref{fig:dioptric},
respectively, show the two possible designs.

In both configurations, a filter blocks the secondary orders and
restricts the RVS wavelength range to the adopted domain:
[848,~874]~nm. The location of the filter has not yet been decided.
For the dioptric option, two locations have been proposed: just before
the grism or just before the CCD plane. The first option would limit
the amount of parasitic light due to secondary reflections on the
filter. In the second case the filter would add more effectively to
the radiation shielding protecting the detectors.

The spectral dispersion is oriented along the scan direction. The
choice of the orientation of the spectra has been partly driven by the
shape of the CCD pixels. Because of manufacturing constraints, the
pixels are rectangular: narrower in the along-scan direction than in
the across-scan direction (see \S~\ref{focal}). Therefore, for the
same resolution $R = 11500$ and sampling (the spectra are sampled at 2
pixels per resolution element), the along-scan orientation of the
spectra requires a smaller dispersive power for the grism than the
across-scan orientation.  In this orientation, the scan law
(\S~\ref{observation}) acts to broaden spectra in the spatial
direction, rather than in the spectral (which would lead to a
reduction in spectral resolution). The along-scan orientation also
ensures that the spectra of stars located close to the upper or lower
boundary of the field of view will not extend beyond the edge of the
CCDs.

\subsection{Focal plane} \label{focal}
\begin{figure}
\resizebox{\hsize}{!}
{\rotatebox{0}{\includegraphics{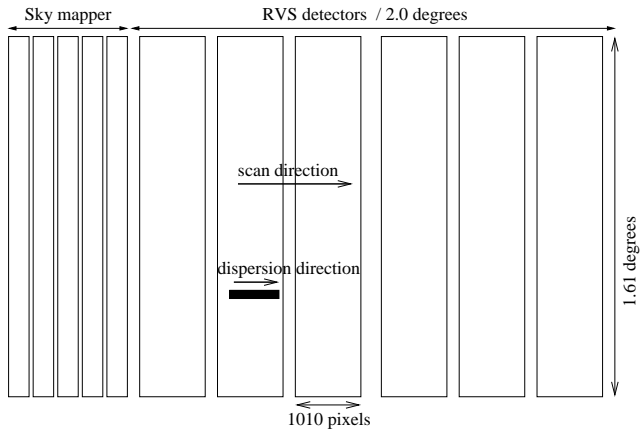}}}
\caption{Schematic view, projected on the sky, of the RVS sky-mapper
(left) and of the RVS detectors (right). In this figure, the 5
sky-mapper CCDs are represented side-by-side. In practice they could
be mixed with other detectors with different functions
(e.g. photometric detectors).}
\label{fig:focalplane}
\end{figure}

The Spectro telescope illuminates three different instruments
located in two distinct physical planes. The RVS sky-mapper (whose role
is described in \S~\ref{mapper}) and the photometric instrument are
located at the telescope focus. The RVS CCDs are located at the
spectrograph focus. As shown in Fig.~\ref{fig:focalplane}, the RVS sky-mapper
(RVSM) field of view precedes that of the RVS detectors 
(with respect to the scan orientation).\\

\subsubsection{Radial Velocity Spectrometer Sky-mapper} \label{mapper}
Unlike {\it Hipparcos}, {\it Gaia} will not rely on an input
catalogue. The satellite will continuously scan the sky and the
sources will be detected on-board in real time by dedicated
instruments. In the case of the RVS, this role will be devolved to the
RVSM. Subsequently, in order to save telemetry, only the RVS pixels
illuminated by a spectrum are transmitted to the Earth.\\

The RVSM is illuminated in undispersed light. In the current
baseline\footnote{The number of RVSM CCDs is still under consideration
and is subject to change.}, it is composed of 5 CCDs. Four CCDs are
illuminated in white light (to maximise the number of photons
collected and increase the detection limit) and one is coated with a
red filter (see below). In order to follow the scanning motion of the
satellite, the 5 CCDs are operated in Time Delay
Integration\footnote{The charges are ``continuously'' transfered from
pixel column to pixel column, in order that they ``follow'' the
sources in their motions across the field of view.} (TDI) mode and
read continuously.  The data from the white-light illuminated CCDs are
analysed on-board in real time by a source detection and centroiding
algorithm. The detection diagnostics coming from these CCDs are
compared in order to reject false detections (e.g. cosmic rays, CCD
defects).  The centroiding information from the validated detections
is used to predict the location and readout times of the spectra on
the RVS detectors.

The wide field of view of the RVSM will allow the detection algorithms
to identify fast moving objects ({\it i.e.} solar system minor bodies and in
particular the Near Earth Objects - NEO) and to derive a first estimate of
their along- and across-scan motions. The observability of NEOs by {\it Gaia} 
is discussed in detail in \citet{mignard2002}.

The fifth CCD will use the same filter as the RVS. This CCD will be
used to derive the magnitudes of the sources in the RVS band and to
map the sky background with a spatial resolution of $\sim30$
arcsec. In fields of high stellar density, the RVS band magnitudes are
required to disentangle overlapped spectra during the subsequent
processing on the ground (see \S~\ref{crowding}).

\subsubsection{Spectrograph focal plane}
As shown in Fig.~\ref{fig:focalplane}, the 2.0 degree width of the RVS
field of view is filled by 6 CCDs\footnote{The number of RVS CCDs is
still under consideration and is subject to change.}. The design of
the RVS focal plane assembly was driven mainly by the issue of the
motion of the spectra across the RVS field of view. Because of the
{\it Gaia} scan law \citep{lindegren1998}, there is a periodic drift
of the spectra in the across-scan direction. The transverse velocity
of the spectra follows a sine law of period 6 hours (the spacecraft
spin period). Its mean semi-amplitude is 0.168~arcsec~s$^{-1}$, but
this semi-amplitude itself varies on a 6 month timescale from 0.163 to
0.174 arcsec~s$^{-1}$ because of the eccentricity of the Earth's
orbit.  This means that there is a spatial smearing of the spectra, up
to $\simeq 20$ arcsec ($\simeq$ 14 pixels) for a transit across the 2
degree width of the RVS focal plane.

Such spatial smearing is prohibitive in terms of loss of
signal-to-noise ratio (since the region from which the spectrum is
extracted will be larger and will hence contain more background) and
substantially enhances the problems arising from overlapping
spectra. Several approaches have been identified and compared to
minimise this smearing.\\

In one approach, a mechanism was envisaged to rotate the detector
focal plane to compensate for the transverse motion. Other items that
could be actuated instead include the dispersing element or any pair
of fold mirrors in the optical path.  Besides the inherent complexity,
the difficulty with this solution is the requirement that the
generation of disturbing torques and variable thermal dissipation
within the satellite (which would particularly affect the astrometric
measurements) should be kept extremely low.

The second approach is to permit the transverse motion, but to split
up the focal plane into smaller detector subunits in the scan
direction. The transverse drift per observation ({\it i.e.} per CCD)
is limited by the shorter exposure time per detector. The total number
of pixels in the focal plane, as well as the total exposure time per
transit (summed over the smaller detector subunits), is kept
unchanged. Assuming 6 CCDs (and 16.5~s crossing time per CCD), the
maximum transverse drift per observation is $\simeq 3$ arcsec ($\simeq
2$ pixels). This is the technique used in the astrometric payload. The
absence of an actuation mechanism is an advantage, but the main
consequence of this solution is an increase in the number of readouts
for a spectral scan, causing reduced performance because of an
increase in the readout noise: the instrument performance is
readout-noise limited. This drawback could be minimised through the
use of innovative detector technologies (L3CCDs) which exhibit very
low readout noise and by careful design of the detector electronics.

The tilt mechanism and the 6 L3CCD configurations yield similar
performances. The second option (6~L3CCDs) was ultimately adopted
because of its lower technical complexity and failure risk. The
characteristics of the RVSM and RVS detectors are summarised in
Table~\ref{tab:focal}.\\

\begin{table}
\caption{Summary of the RVSM and RVS characteristics. The CCD dimensions
and total noise are given for one CCD. The overall efficiency
of the RVSM is given for the CCD coated with the RVS filter. The overall
efficiency includes the telescope reflectivity, the optics (RVS only), grism
(RVS only) and filter transmissions and the CCD quantum efficiency (QE).}
\begin{center}
\begin{tabular}{l c c} \hline
                              & RVSM               & RVS \\ \hline
Type of CCD                   & ``classical''      & L3 CCD             \\
Operating mode                & TDI                & TDI                \\
Number of CCDs                & 5                  & 6                  \\
CCD dimensions (pixels)       & $336 \times 3930$  & $1010 \times 3930$ \\
CCD dimensions (degrees)      & $0.09 \times 1.61$ & $0.28 \times 1.61$ \\
Exposure time /CCD (s)        & 5.5                & 16.5               \\
Pixel size (microns)          & $10 \times 15$     & $10 \times 15$     \\
Pixel size (arcsec)           & $0.98 \times 1.47$ & $0.98 \times 1.47$ \\
CCD total noise (e$^-$/pixel) & 9                  & 2                  \\
CCD QE ($\lambda=500$ nm)             & 66\%               & $--$       \\
CCD QE ($\lambda=700$ nm)             & 92\%               & $--$       \\
CCD QE ($\lambda=860$ nm)             & 71\%               & 71\%       \\
Filter transmission                   & 85\%               & 85\%       \\
Overall efficiency ($\lambda=860$ nm) & 40\%               & 21\%       \\
\end{tabular}
\end{center}
\label{tab:focal}
\end{table}

One further possible solution is elegant but presently unproven in the
context of the RVS. This would be to move the signal ``diagonally''
within the detector by means of specialised structures on the CCD
itself (2-dimensional clocking), thus compensating for the across-scan
drift. This solution could also be effective in compensating for the
optical distortion.  While this technology still needs to be validated
for space applications, it has begun to be used in ground observations
\citep{howell2003}.  The use of 2D clocking CCDs for the RVS
instrument is under evaluation. This solution will be reconsidered for
inclusion in the baseline at the end of phase B1 (end of first quarter
2005).

\begin{figure*}
\begin{center}
\resizebox{\hsize}{!}{
\includegraphics{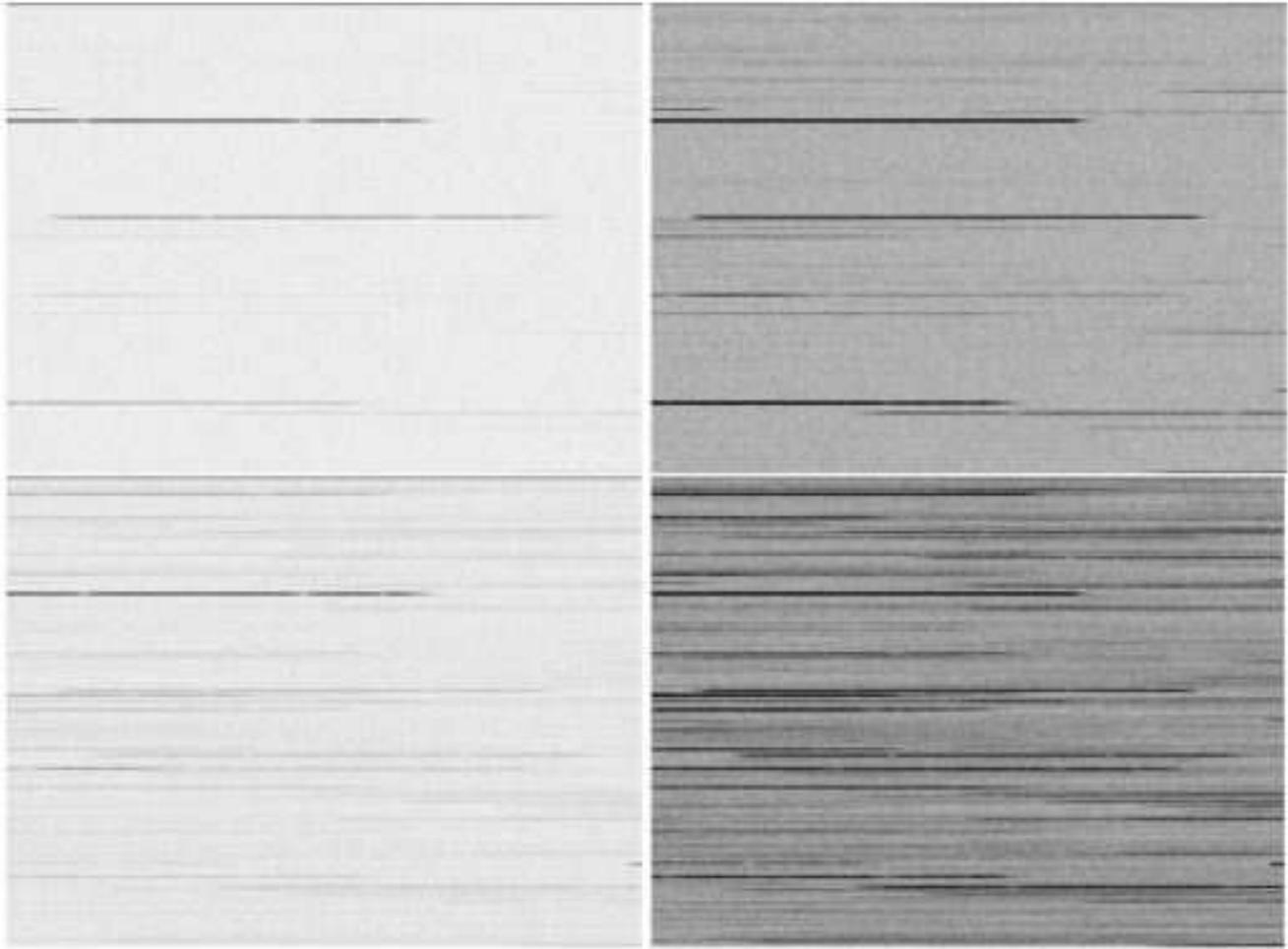}}
\end{center}
\caption{Simulated {\it Gaia}-RVS images for high Galactic latitudes
(top) and $|b|\leq5^{\circ}$ (bottom), in linear scale (left) and
in ``thresholded'' linear scale (right). The images are generated for
1 transit of the focal plane of 6 CCDs, each with a readout noise of
2e$^{-}$.}
\label{fig:sim_image}
\end{figure*}

\subsubsection{RVS observations}
The RVS will be operated in TDI scan mode. Therefore, the raw RVS
observations will appear as a ``never-ending'' strip of (spectrally)
dispersed sky. However, in order to save on the telemetry budget, only
the parts of the raw observations containing spectra will be
downloaded to the Earth.

Fig.~\ref{fig:sim_image} shows snapshots of simulated raw RVS
observations for high (top) and low Galactic latitudes (bottom).
The images on the left and on the right contain the same stars and
only differ by the luminosity scales used to code the number of
photo-electrons recorded per pixel (and per transit). The left-hand
side images are coded with an inverse linear luminosity scale.
In the right-hand side images, the pixels whose intensities are in
the range $[1, 125]$ e$^-$ are coded with an inverse linear
luminosity scale and the pixels which exceed 125 e$^-$ are coded
in black. 

These images have been generated taking into account the main RVS
characteristics: spectral resolution, instrumental throughput
efficiency, filter function, spatial and spectral
point-spread-function (PSF), detector pixel
dimensions, photon noise, readout noise, a model for the Zodiacal
light contribution and the spatial smearing due to the scan law. The
images have been generated down to $V=20$ using the star density
counts as function of Galactic latitude from Table~6.6 of the
\citet{esa2000} report and the distribution of spectral types as given
in their Table~6.4. The star positions have been generated randomly
for this assumed stellar density. Spectra of the appropriate
brightness and spectral type have been simulated by scaling model
spectra taken from the library of \citet{zwitter2003lib, zwitter2004}.
The RVS field of view simulator is accessible and can be run on-line
via the simulator web site:
http://www.mssl.ucl.ac.uk/gaia-rvs/simulator.html.

What is immediately clear from the simulated images in
Fig.~\ref{fig:sim_image} is the effect of the overlapping of spectra
in low Galactic latitude fields. Less evident, but partly visible in
the right-hand side images, is the number of spectra at low light levels
in these images (at the $V=17.5$ limit, spectra have accumulated less
than 1 photon per pixel per transit). Indeed, the spectra visible on the
right-hand side of Fig.~\ref{fig:sim_image} are all from stars with
$V \leq 15$. This emphasises the extent to
which the readout noise (2 e$^-$ per pixel per CCD) dominates the signal
at the limiting magnitude, except in the Galactic plane where the
dominant background is due to faint stars. Therefore, the derivation of radial
velocities at the limiting magnitude will be possible only after the
co-adding of all of the scans of a given target taken by the end of
the mission.  This has critical implications for the data-handling,
compression and transmission and rules out the possibility of using
lossy compression algorithms in transferring the data to the ground.

In Fig.~\ref{fig:sim_image}, the background light (here only zodiacal
light) displays no spectral signature. This will be the case in all
the regions of the sky where the background light surface brightness
is uniform on an angular scale of a few arcminutes: the spectral
information will be smoothed by the slow variation of this source over
the field of view.

The simulated images differ from the actual RVS raw observations in
that they do not yet include the effects of pixel to pixel variability
within the detector, any cosmetic defects in the detectors, or cosmic
rays. In addition, they do not include the ``small'' effect of
spectral and spatial smearing due to the ``small'' distortions from
the telescope plus spectrograph optics, nor the effects of co-adding
the signal from several detectors that might not be perfectly
aligned. In addition, the spectral catalogue contains only ``normal'',
single stars.\\

\subsection{Calibration strategies} \label{calibration}
The RVS, as well as the other {\it Gaia} instruments, presents several
peculiarities with respect to classical ground-based instruments: it
will scan the sky continuously with (hopefully) no interruption during
the 5 years of the mission; it will observe $\sim 100-150\times10^{6}$
stars; each star will be observed a large number of times; the
instrument will be extremely optically and mechanically stable. All
these items support the idea that the RVS instrument could be
self-calibrated using a global iterative approach.

The principle of the global iterative calibration is to use an
iteratively self-selected sample of ``well-behaved'' stars ({\it i.e.}
sufficiently bright; astrometrically, photometrically and
kinematically stable over successive observations; of a spectral type
well-suited for the wavelength calibration) to derive, iteratively,
the radial velocities of the iteratively selected stars, calibrate the
wavelength scale and calibrate the instrument response and geometry.

The calibration process will use the information provided by the other
{\it Gaia} instruments: very accurate source positions, very good
knowledge of the satellite attitude and of the magnitudes, colours and
atmospheric parameters of the sources. Pre-flight calibrations, as
well as ground-based radial velocity standards (selected, for example,
from among the Solar-type stars surveyed for extra-solar planets),
will be used to initiate the iterative process. With the progress of
the mission, {\it Gaia} will iteratively select thousands of
appropriate standards, which will complement the ground-based
standards and will provide a very constrained grid in the spatial and
temporal dimensions. Finally, the ground-based standards will fix the
zero point of the radial velocities.

The full-precision results will be obtained after completion of the
observations and after convergence of the iterative auto-calibration
procedure. The final {\it Gaia} catalogue should be available a few
years after mission end: {\it i.e.} around 2017-2020. Intermediate
releases, of lower precision than the final catalogue, could be
envisaged a few years after launch: e.g. $\sim$~2012-2014.

The procedure described above is presently the baseline for the
calibration of the RVS and in particular its wavelength calibration.
Studies are also in progress to assess whether complementary
techniques could be used. The slit-less, objective-prism-like modus
operandi of the {\it Gaia} spectrograph prevents the use of standard
ground-based calibration techniques, involving lamps or sky lines. The
only possibility to {\em mark} wavelengths on {\it Gaia} spectra is to
absorb selective wavelengths. Should it emerge during the RVS final
system design that it would be a useful tool to implement, an
absorption cell could be inserted in the optical chain, filled with an
appropriate absorbing medium able to mark strong and widely spaced
absorption lines away from those of major astrophysical
relevance. Efforts to identify an absorbing medium appropriate for the
{\it Gaia} operating conditions are underway \citep{desidera2003,
pernechele2003}.  Alternatively, a bundle of multi-mode optical fibres
carrying suitable Bragg gratings, inserted in the collimated beam,
could be considered. These alternative ways of wavelength calibrating
{\it Gaia} spectra would probably be of greater interest in the pre-,
during- and post-flight ground-based observations, when the
maximisation of the science exposure time and the increase of the
accuracy are particularly sought after. Particularly promising in this
respect are the fibre Bragg gratings that are suitable for mass
production.

\subsection{Observation strategy} \label{observation}
The Gaia satellite will scan the sky continuously during the 5 years
of the mission. As in the case of its precursor {\it Hipparcos}, a
uniform revolving scan has been proposed for {\it Gaia} and parameters
for the scanning law were calculated to produce a smooth sky coverage
and achieve the best accuracy on the measurements of the astrometric
parameters.

The {\it Gaia} Nominal Scanning Law (NSL) is described in
\citet{lindegren1998}.  Some of its parameters have been modified
(Spring 2002) to take into account changes in the payload design due
to the change of launcher from Ariane V to Soyouz-Fregat.

{\it Gaia} will rotate around its z-axis with a uniform velocity of
$\omega = 60$ arcsec s$^{-1}$ so that each instrument will scan a
great circle on the sky in 6 hours. The rotational axis itself
precesses at a constant angle of 50 $\deg$ around the direction from
the satellite to the centre of the Sun.  The mean speed of the
rotational z-axis on the sky, in units of the apparent Sun speed, is
equal to 4.095; this leads to a mean number of 5.200 (overlapping)
precession loops of the z-axis on the sky per year, each loop being
covered in $\sim 70$ days. A complete sky coverage is achieved in one
year through motion of the axis of the precession cone (which points
towards the Solar centre) along the ecliptic.

\begin{figure}
\resizebox{\hsize}{!}{
\includegraphics{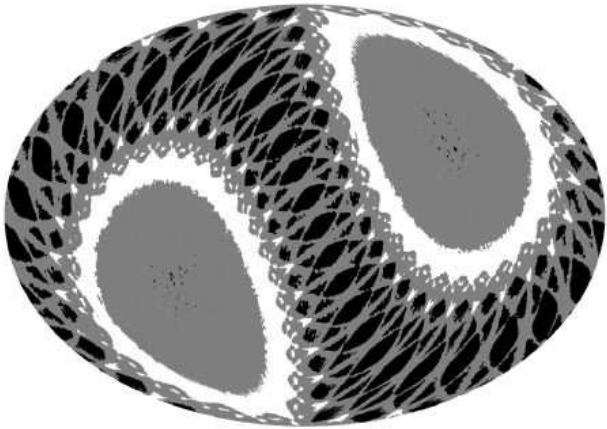}}
\caption{Number of transits through the RVS-field of view versus
Galactic coordinates for the 1800 day {\it Gaia} mission. There are
three ``grey'' levels: $\leq$ 80 transits (black), between 80 and 120
transits (grey) and $\geq$ 120 transits (white).}
\label{fig_transits_gal}
\end{figure}

The NSL makes it possible to predict the attitude of the satellite as
a function of time throughout the mission. Using the attitude
parametrisation of {\it Gaia}-NSL based on the quaternion formalism
derived by \citet{lindegren2000} and the FORTRAN code he provided
\citep{lindegren2001}, the total number of transits during the assumed
5 year mission (arbitrarily starting from J2000) was computed as
function of position on the sky for integer values of Galactic
longitude and latitude. The number of transits versus Galactic
coordinates is plotted in Fig.~\ref{fig_transits_gal}.  A similar
figure, showing the number of RVS transits as function of ecliptic
coordinates, is presented by \citet{pourbaix2003}.  According to the
position on the celestial sphere, the number of transits for the RVS
instrument ranges from 54 to 239, with a mean value (averaged over the
sky) of~102.

The RVS will provide radial velocities for late type stars up to
magnitude $V \simeq 17-18$ (see \S~\ref{radial}). The star count model
of \citet{chen1997} has been used by \citet{torra1999} (updated values
are in given in \citeauthor{esa2000} \citeyear{esa2000}) to evaluate
the number of stars to be observed by {\it Gaia} as function of
magnitude, stellar population, stellar type and Galactic
coordinates. The model predicts that 100-150 million stars will be
observed up to magnitude $V =$~17-18. Approximately 80\% of those
stars will belong to the thin disk, $\simeq$~15\% to the thick disk
and $\simeq5-10$\% to the halo and the bulge. The majority ($\simeq
70$\%) of RVS targets will be G and K type stars.

\begin{figure*}
\resizebox{\hsize}{!}{\includegraphics{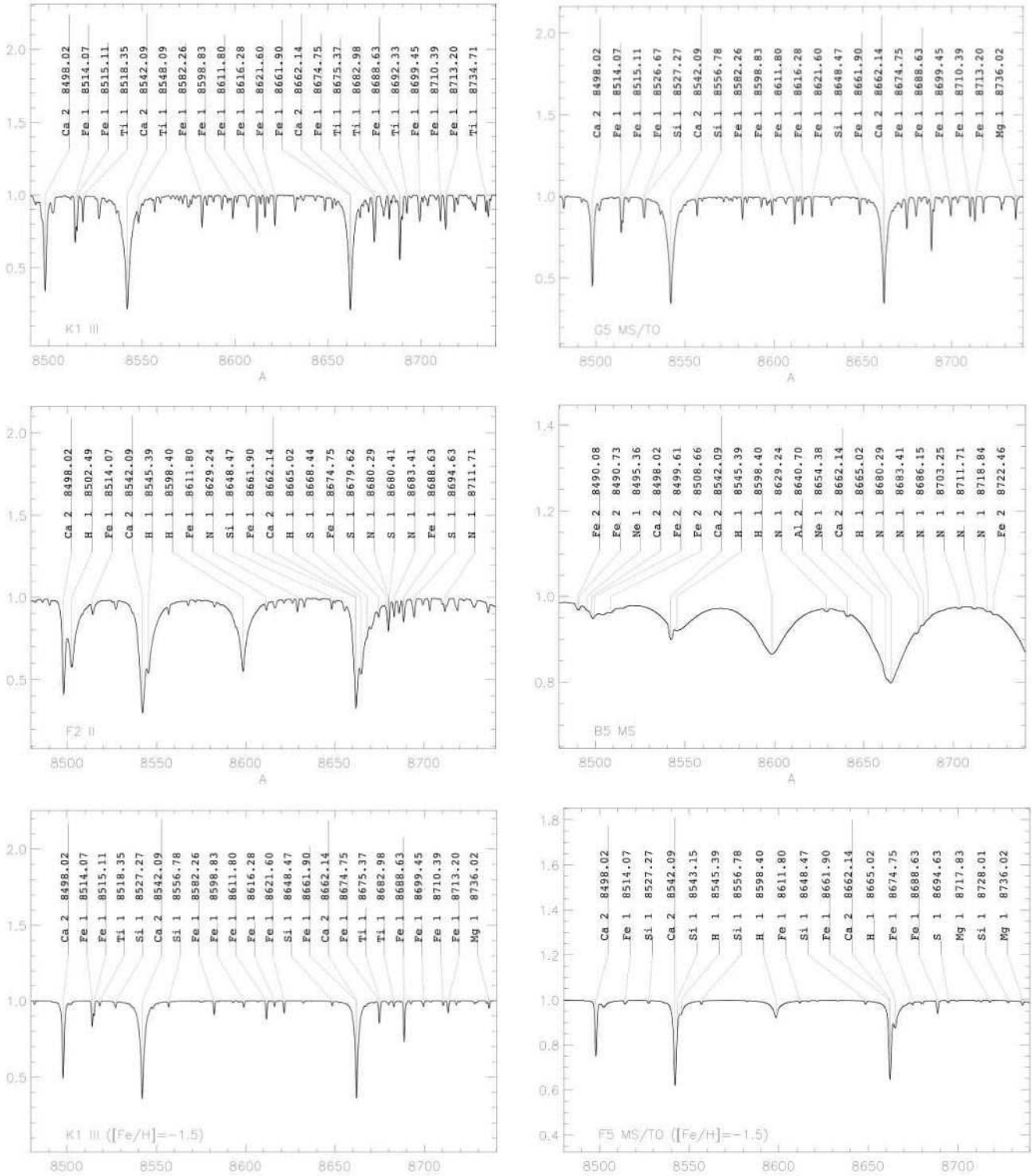}}
\caption{Six examples of RVS-like spectra. The panels show 4 solar
metallicity stars: G5 turn-off (top left), K1 III (top right), F2 II
(middle left), B5 main sequence (middle right); and two metal-poor
stars: F5 turn-off (bottom left) and K1 III (bottom right). See
Table~\ref{tab:tracers} for details.}
\label{fig:spectra}
\end{figure*}

\section{Stellar and interstellar parameters} \label{stellar}
\subsection{Spectra} \label{spectra}
In the RVS wavelength range, [848, 874] nm, the spectra of F, G and K
type stars are dominated by the ionised Calcium triplet (849.80~nm,
854.21~nm, 866.21~nm).  Although the strength of the triplet lines
decreases with increasing gravity, it nevertheless remains very strong
in dwarf stars: their equivalent widths are larger than 0.3~nm for
stars of spectral type F8V to M6V.  The triplet is non-resonant and
therefore will not be affected by contributions from the interstellar
medium. In addition to the Calcium triplet, the RVS spectra of
late-type stars will also contain numerous unblended weak lines of
astrophysical interest and in particular alpha-elements like Si~I or
Mg~I. In the coolest stars, CN and TiO molecular transitions are
visible, but no strong molecular bandhead is present.

In early-type stars, the strongest features are the Paschen lines
whose equivalent widths are strongly anti-correlated with gravity. The
spectra of hot stars also exhibit some other strong (e.g. Ca~II, N~I)
and weak (e.g. He~I, He~II) lines.

In addition to the stellar features, the RVS wavelength range contains
a diffuse interstellar band (862.0~nm) which appears to be a reliable
tracer of interstellar reddening (see \S~\ref{interstellar}).

Fig.~\ref{fig:spectra} shows examples of synthetic RVS-like spectra
for 6 representative tracers of Galactic structure: a K1 III star and
a G5 main sequence/turn-off (MS/TO) star of solar metallicity (typical
of the disk); an F2 supergiant and a B5 MS star (tracers of spiral
structure); a K1 III star and an F5 MS/TO star, both metal-poor
(typical of the halo).  The parameters of the 6 stars are summarised
in Table~\ref{tab:tracers}.  They have been computed using the VALD
atomic data \citep{pi95, ryabchikova1999, ku99}, the Kurucz
atmospheric model \citep{ku93cd13} and N. Piskunov's SYNTH program
\citep{piskunov1992}.  All models and spectra have been computed
assuming a micro-turbulence velocity of 2 km~s$^{-1}$ and
plane-parallel static atmospheres.\\

\begin{table}
\caption{Parameters of the 6 representative tracers of Galactic
structure plotted in Figure~\ref{fig:spectra}.}
\begin{center}
\begin{tabular}{l r r r r r} \hline
         & Teff  & $\log$ g & [Fe/H] & [$\alpha$/Fe] & $v \sin i$ \\ \hline
K1 III   &  4500 & 2.0      &  0.0   &  0.0          & 5.0        \\
G5 MS/TO &  5500 & 4.0      &  0.0   &  0.0          & 5.0        \\
F2 II    &  7000 & 1.0      &  0.0   &  0.0          & 20.0       \\
B5 MS    & 15000 & 4.5      &  0.0   &  0.0          & 50.0       \\
K1 III   &  4500 & 2.0      & -1.5   & $+$0.4        & 5.0        \\
F5 MS/TO &  6500 & 4.0      & -1.5   & $+$0.4        & 20.0       \\
\end{tabular}
\label{tab:tracers}
\end{center}
\end{table}

The easy accessibility from the ground of the {\it Gaia} wavelength
range was one of the factors emphasised by \citet{munari1999} when
proposing the $[848-874]$~nm interval, which was at that time also
discussed within the GAIA Science Advisory Group as one of the possible 
wavelength ranges for the mission's spectrograph. This interval is also
remarkably clear of telluric absorption, unlike the rest of the
$\lambda \geq$ 670~nm far-red optical region.

This allows an intensive observational campaign to be conducted with
ground-based telescopes with the aim of deepening our knowledge of
this wavelength interval at spectral resolutions equal to, or larger
than, the 11\,500 foreseen for {\it Gaia}. By the time that {\it Gaia}
will start to deliver its spectral data, it is to be anticipated that
it will become the reference interval for classification and
analysis. A wide range of spectroscopic follow-up options will be open
to ground-based telescopes, including either a similar observing
strategy to that of {\it Gaia} but to fainter limits, or the extension
of the time coverage of variable phenomena.

Inspection of the Asiago Database of Spectroscopic Databases (ADSD)
(\citealt{sordo2003}; Sordo \& Munari 2004, in preparation), listing
259 atlases for the optical region, returns 12 spectral atlases
containing more than 15 objects over the {\it Gaia} wavelength
interval at a resolution larger than 3\,000. These 12 atlases offer
some insight into the appearance of real stars in this wavelength
region, including peculiar types, and the cooler spectral types
(F,G,K) that will dominate the field star population at the {\it Gaia}
magnitudes are particularly abundant. The characteristics of the 12
atlases are summarised in Table~\ref{tab:atlases}.

No special observational effort appears to be required to assemble new
spectral atlases surveying normal stars over the HR diagram in this
spectral region. However, more work on metal-poor, peculiar and
molecular-band stars (M, C, S) can and should be carried out.

\begin{table}
\caption{Spectral atlases with a minimum sample of 15 stars and a
resolving power larger than 3\,000 over the {\it Gaia} 848--874~nm
spectral range.  The columns give the resolving power (R.P.),
the number of stars included and the range of their spectral types.}
\label{tab:atlases}
\begin{center}
\begin{tabular}{l r r c} \hline
atlas                       & R.P.  & N   & spec.     \\
                            &       &     & type      \\ \hline
\citet{andrillat1995}       &  3000 &  76 & O5-G0,pec \\
\citet{cenarro2001}         &  5000 & 706 & O6-M8     \\
\citet{carquillat1997}      &  6500 &  54 & A2-M4,C,S \\
\citet{valdes2004}          &  7200 &1273 & A0-M8,C,S \\
\citet{serote1996}          &  8000 &  21 & B3-M5     \\
\citet{fluks1994}           &  8600 &  22 & MIII      \\
\citet{montes1999}          & 13000 & 132 & F0-M8     \\
\citet{munaritomasella1999} & 20000 & 130 & O4-M8     \\
\citet{marrese2003}         & 20000 &  94 & F2-M7     \\
\citet{pavlenko2003}        & 20000 &  24 & Carbon    \\
\citet{munari2003}          & 20000 &  17 & pec       \\
\citet{montes1998}          & 55000 &  48 & F5-M8     \\
\end{tabular}
\end{center}
\end{table}

\begin{table*}
\caption{Precision (one sigma error) of the RVS instrument in the
derivation of radial velocities as a function of magnitude, for a
single observation (top) and at the end of the mission (102
observations: bottom). The performances are expressed in km~s$^{-1}$.}
\label{tab:performances}
\begin{center}
\begin{tabular}{l c c | r r r r r r r r r r}
         &     &               & \multicolumn{10}{c}{Single transit}                                                     \\ \hline
         & $[Fe/H]$ & $v \sin i$    & \multicolumn{10}{c}{V}                                                                  \\
         &      & (km~s$^{-1}$) & 9        & 10       & 11       & 12     & 13     & 13.5   & 14     & 14.5   & 15     & 15.5    \\ \hline
K1 III   & 0.0  & 5           & $\leq$ 1 & $\leq$ 1 & $\leq$ 1 & 1.1    & 2.3    & 3.1    & 4.7    &  7.2   & 11.5   & 34.9   \\
G5 MS/TO & 0.0  & 5           & $\leq$ 1 & $\leq$ 1 & 1.2      & 2.1    & 4.4    & 6.3    & 9.7    & 19.7   & $>$ 35 & $>$ 35 \\
F2 II    & 0.0  & 20          & $\leq$ 1 & 1.3      & 2.3      & 4.3    & 9.3    & 14.4   & 24.3   & $>$ 35 & $>$ 35 & $>$ 35 \\
F2 II    & 0.0  & 50          & 1.3      & 2.1      & 3.6      & 6.5    & 13.8   & 21.2   & 33.4   & $>$ 35 & $>$ 35 & $>$ 35 \\
B5 MS    & 0.0  & 50          & 14.8     & 25.0     & $>$ 35   & $>$ 35 & $>$ 35 & $>$ 35 & $>$ 35 & $>$ 35 & $>$ 35 & $>$ 35 \\
B5 MS    & 0.0  & 150         & 18.3     & 31.1     & $>$ 35   & $>$ 35 & $>$ 35 & $>$ 35 & $>$ 35 & $>$ 35 & $>$ 35 & $>$ 35 \\
K1 III   & -1.5 & 5           & $\leq$ 1 & $\leq$ 1 & 1.1      & 1.8    & 3.6    & 5.0    & 7.5    & 11.7   & 26.8   & $>$ 35 \\
F5 MS/TO & -1.5 & 20          & 1.2      & 1.9      & 3.2      & 5.8    & 12.3   & 24.3   & $>$ 35 & $>$ 35 & $>$ 35 & $>$ 35 \\
F5 MS/TO & -1.5 & 50          & 2.1      & 3.4      & 5.6      & 10.7   & 22.3   & $>$ 35 & $>$ 35 & $>$ 35 & $>$ 35 & $>$ 35 \\
\end{tabular}
\vskip 0.5 cm

\begin{tabular}{l c c | r r r r r r r r r r}
         &   &               & \multicolumn{10}{c}{Mission (102 transits - average case)}                           \\ \hline
         & $[Fe/H]$ & $v \sin i$    & \multicolumn{10}{c}{V}                                                               \\
         &   & (km~s$^{-1}$) & 12       & 13       & 14       & 15     & 15.5   & 16     & 16.5   & 17     & 17.5   & 18     \\ \hline
K1 III   & 0.0 & 5           & $\leq$ 1 & $\leq$ 1 & $\leq$ 1 & 1.1    & 1.6    & 2.6    & 3.8    &  6.2   & 10.6   & 25.5   \\
G5 MS/TO & 0.0 & 5           & $\leq$ 1 & $\leq$ 1 & $\leq$ 1 & 2.1    & 3.4    & 5.0    & 8.3    & 16.0   & $>$ 35 & $>$ 35 \\
F2 II    & 0.0 & 20          & $\leq$ 1 & $\leq$ 1 & 2.0      & 4.8    & 7.6    & 12.6   & 20.8   & $>$ 35 & $>$ 35 & $>$ 35 \\
F2 II    & 0.0 & 50          & $\leq$ 1 & 1.3      & 3.1      & 7.5    & 11.4   & 18.0   & 29.9   & $>$ 35 & $>$ 35 & $>$ 35 \\
B5 MS    & 0.0 & 50          & 8.2      & 18.6     & $>$ 35   & $>$ 35 & $>$ 35 & $>$ 35 & $>$ 35 & $>$ 35 & $>$ 35 & $>$ 35 \\
B5 MS    & 0.0 & 150         & 10.0     & 23.5     & $>$ 35   & $>$ 35 & $>$ 35 & $>$ 35 & $>$ 35 & $>$ 35 & $>$ 35 & $>$ 35 \\
K1 III   & -1.5 & 5          & $\leq$ 1 & $\leq$ 1 & $\leq$ 1 & 1.7    & 2.5    & 4.1    & 6.3    & 10.3   & 23.4   & $>$ 35 \\
F5 MS/TO & -1.5 & 20         & $\leq$ 1 & 1.1      & 2.6      & 6.1    & 9.5    & 17.3   & $>$ 35 & $>$ 35 & $>$ 35 & $>$ 35 \\
F5 MS/TO & -1.5 & 50         & 1.0      & 2.0      & 4.4      & 10.5   & 16.9   & 31.6   & $>$ 35 & $>$ 35 & $>$ 35 & $>$ 35 \\
\end{tabular}
\end{center}
\end{table*}

\subsection{Radial velocities} \label{radial}
\subsubsection{Performances}

The initial motivation for the implementation of a spectrograph on
board {\it Gaia} was the acquisition of radial velocities.  Their
determination has, to a large extent, driven the definition and
optimisation of the characteristics of the RVS. Throughout the
definition phase, studies were conducted to evaluate the performance
of the different proposed configurations. Those studies all rely on
similar Monte-Carlo approaches to derive the RVS radial velocity
precision: cross-correlating a large number of RVS-like spectra,
either observed \citep{munari2001, david2001} or synthetic
\citep{katz2000, katz2002, david2002a, david2002b, zwitter2002,
munarietal2003} against suitable templates.

Subsequently, the simulations have been re-run to evaluate the
performance of the RVS adopted baseline as a function of magnitude
and for the 6 representative tracer populations listed in
Table~\ref{tab:tracers}. In addition, for each of the three spectral
types, F2 II, B5 MS and F5 MS/TO, which could present high rotational
velocities, two projected rotational velocities have been considered:
a ``low'' and a more ``typical'' value (see
Table~\ref{tab:performances}).

To generate RVS-like observations, \citet{ku93cd18} synthetic spectra
have been convolved to the RVS spectral resolution, normalised to the
star magnitude (according to its atmospheric parameters), sampled
(assuming two pixels per PSF full width at half maximum - FWHM) and
degraded with photon, background and detector noise, using a simulator
of the RVS instrument (Katz 2004, in preparation). The simulations
took into account zodiacal light and faint background stars (assuming
a total surface brightness of $V=21.75$ mag~arcsec$^{-2}$), the
telescope pupil area, the overall instrument efficiency, the exposure
time, the spectral profile perpendicular to the dispersion direction
and detector noise.

The performances were derived by Monte-Carlo simulations. For each
spectral type and magnitude, 1000 or 2000 cross-correlations of
RVS-like spectra with Kurucz synthetic templates (convolved at the RVS
resolving power) were performed in direct space. In each case, the
precision was derived using a robust estimator of the standard
deviation of the distribution of (1000 or 2000) estimated radial
velocities. After sorting the radial velocities, the standard
deviation was estimated using the following formula:
\begin{equation}
\sigma = \frac{V_{R(0.8415)} - V_{R(0.1585)}}{2}.
\end{equation}
Here $V_{R(0.1585)}$ is the 15.85$^{th}$ percentile of the sorted
distribution of radial velocities and $V_{R(0.8415)}$ is the
84.15$^{th}$ percentile of the same distribution.  This robust
estimator gives low weight to the outliers which appear at the
faintest magnitudes (when secondary correlation peaks could be
mistaken for the correct correlation peak). Future work will focus on
the development of techniques to identify and reject these
outliers. The fraction of outliers ({\it i.e.} the fraction of
estimated radial velocities beyond 3 times the estimated dispersion)
is, to a first approximation, a function of the estimated dispersion
and of the stellar type. For $\sigma =$~5 km~s$^{-1}$ there are
$\simeq$1.5\% and $\simeq$0\% outliers, respectively, in the cases of
a G5V star and of a F2 II star. The fractions of outliers become
$\simeq$12\% and $\simeq$7\% respectively for $\sigma =$~20
km~s$^{-1}$.

In these simulations, the distributions of estimated radial velocities
exhibit no statistically significant bias.

Many small effects still remain to be taken into account in the full
assessment of the RVS accuracy budget. These include optical
aberration, CCD charge transfer inefficiency, instrument and
wavelength calibration. In order to account for these missing effects,
a 40\% explicit error margin was added to the RVS performances. The
radial velocity performances, for single observations and at the end
of the mission (after combining the 102 observations collected over
the 5 years of the mission), are presented in
Table~\ref{tab:performances}. It is likely that the second order
effects will become very significant or even dominant for sufficiently
bright stars, for which the present simulations give 1 km~s$^{-1}$ or
better performance. For this reason, no values below $\sigma
=$~1~km~s$^{-1}$ are given in Table~\ref{tab:performances}.  Above
$\sigma \simeq$~30-40~km~s$^{-1}$ the performance degrades very
rapidly as function of magnitude and the estimation of the performance
becomes imprecise. As a consequence, no values above $\sigma
=$~35~km~s$^{-1}$ are given in Table~\ref{tab:performances}.

\begin{figure*}
\resizebox{\hsize}{!}{
\rotatebox{-90}{\includegraphics{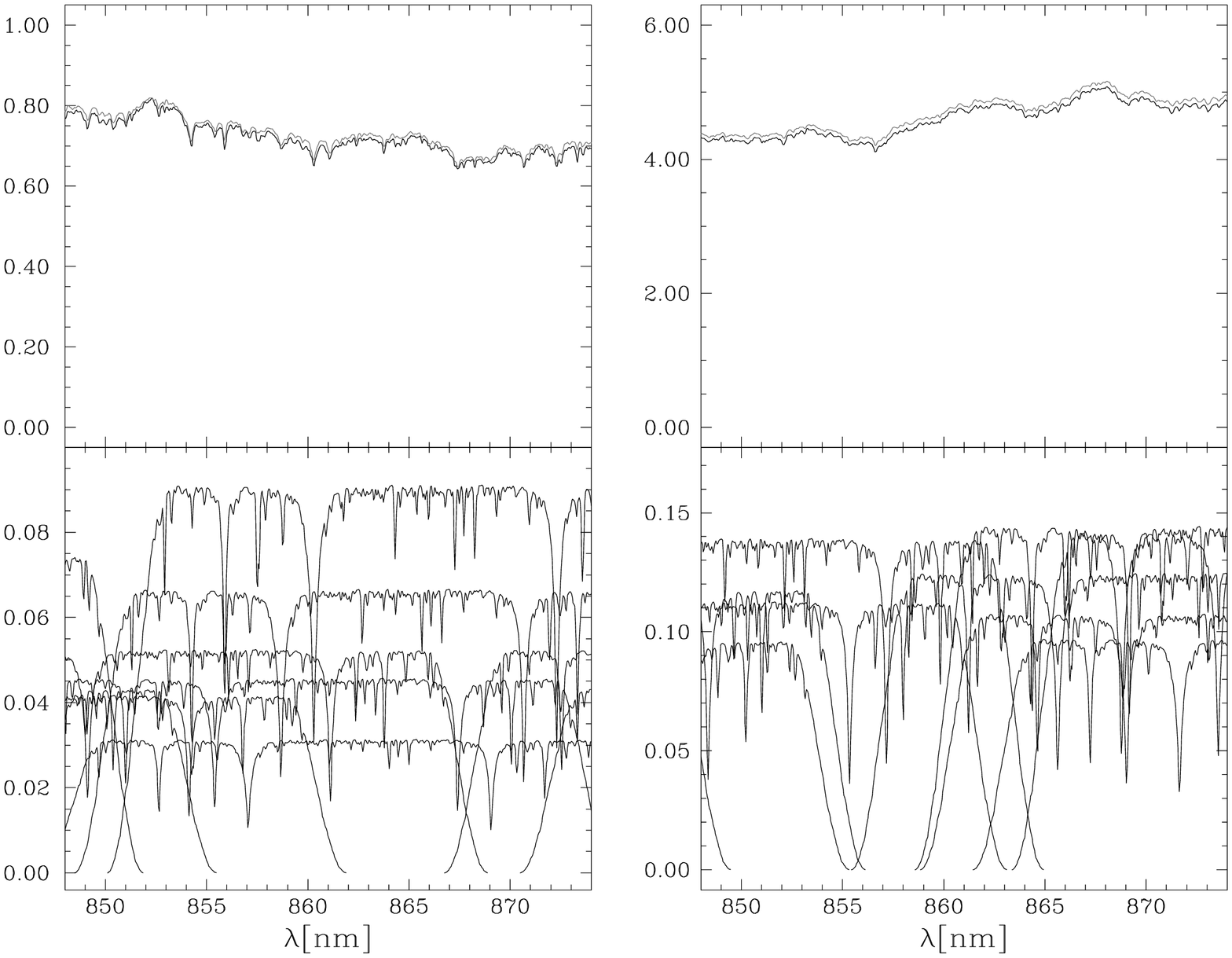}}}
\caption{Crowding of stellar spectra in the focal plane of the
slitless spectrograph results in spectral overlaps. The left panels
depict the typical situation at ``high'' Galactic latitudes: 1000
stars$/$deg$^2$ with $V<17$. The right panels are for a more crowded
region near the Galactic plane: 6000 stars$/$deg$^2$ with $V<17$. The
top panels show the background accumulated over the whole mission (for
a spectral tracing width of 2 pixels) if the transits with overlaps of
the stars brighter than $V=14$ are omitted. The true background is
plotted in black (lower line) and the one recovered from the
astrometric and photometric information from other instruments aboard
{\it Gaia} in grey (upper line).  The bottom panels show, in the same
ordinate units as in the upper ones, the contribution to the
background of the ten brightest stars fainter than V=14. Ordinates are
fluxes in units of the flux from a $V=17$ K1~V star integrated over
the whole mission.}
\label{fig:crowding}
\end{figure*}

The precisions given in Table~\ref{tab:performances} have been
evaluated for the average number of observations per star, namely 102
transits. As described in \S~\ref{observation}, the number of epochs
of observation will vary with Galactic coordinates. As a consequence,
the end of mission performances and the range of magnitude probed by
the RVS will also vary with Galactic coordinates. As an example, for a
G5 MS/TO and for the minimum (54 transits), average (102 transits) and
maximum (239 transits) number of observations, a precision of $\sigma
\simeq 16$~km~s$^{-1}$ is obtained respectively for $V \simeq 16.65$,
$V \simeq 17$ and $V \simeq 17.45$ mag.

Another parameter which varies with Galactic coordinates and which
will impact on the RVS performance is the stellar density. This is
discussed in the next section.

\subsubsection{Crowded regions} \label{crowding}
Because the RVS is an integral field spectrograph, some degree of
spectral overlap is unavoidable, even in regions with a low stellar
density. Consequently, the background to be subtracted from any
spectrum consists not only of the smooth zodiacal light contribution
but also of discrete jumpy spectral tracings of (partially)
overlapping stars. Fig.~\ref{fig:crowding} depicts examples of such
background spectra: on the left, a low density region (1000
stars$/$deg$^2$ with $V<17$) typical for the ``high'' Galactic
latitudes ($|b| \geq 30^\circ$) and on the right, a higher density
region (6000 stars$/$deg$^2$ with $V<17$) representative of the range
of latitudes: $5^\circ \leq |b| \leq 10^\circ$.

Overlaps of the brightest background stars are rare and occur only on
a small number of transits. Their contribution to the overall
background is not, however, negligible and their spectral lines can
degrade the accuracies of the radial velocity or other parameters.
Fig.~\ref{fig:crowding} shows the background accumulated over the
whole mission, excluding transits with overlapping stars brighter than
$V=14$.  The loss of integrated signal corresponding to the rejection
of the transits containing ``bright'' contaminating stars is
acceptably low. Bright ($V<14$) overlappers are present in $\sim 3$\%\
(Galactic halo) or $\sim 13$\%\ (Galactic plane) of all transits. The
bottom panels of Fig.~\ref{fig:crowding} plot individual contributions
from the ten strongest overlapping stars fainter than $V=14$ while the
top panels show the background signal which is the sum of these 10 and
several fainter contributors.  The fluxes are given in units of the
flux from a V=17 K1~V star integrated over the whole mission (102
transits). Note that the accumulated background is very smooth. Lines
from individual overlapping stars become diluted by the continuum
contributions of other overlapping stars in the same and other
transits.

The background can be very well modelled from information that is
available from {\it Gaia}'s astrometric and photometric measurements
(stellar positions, magnitudes, rough spectral types and velocities of
overlapping stars).  In Fig.~\ref{fig:crowding} it is assumed that
stellar positions and magnitudes are accurately known which is
certainly true given the excellent astrometric accuracy combined with
the photometric and star mapper measurements. Also, spectral
information could be derived from photometric observations: in
Fig.~\ref{fig:crowding} a mismatch of 250~K in temperature and 0.5~dex
in metallicity and gravity are assumed. We note that this is
conservative, as the final errors on these parameters are expected to
be roughly half these values.  If, in addition, some knowledge
(provided by the RVS) on the radial velocities of the overlapping
stars is assumed the overall background can be modelled.

From the above paragraph, it would appear that the radial velocities
are simultaneously necessary to model both the background and the
unknown information one wants to extract from the spectra. One
possible approach to solve this problem is to proceed in an iterative
way.  At each iteration, the ``stacked'' spectra are analysed, one by
one, from the brightest to the faintest star.  The analysis of each
spectrum is made of two consecutive steps: modelling and subtraction
of the background and then derivation of the radial velocity. On the
first iteration, the background of a given star is modelled using the
first estimates of the radial velocities of the sources brighter than
this star (which have already been analysed) and using only the level
and shape of the continua of the fainter stars (whose radial
velocities have not yet been estimated). At the end of the first
iteration, an estimate of the radial velocity of each of the "stacked"
stars has been derived. The process is then iterated. The new or
refined estimates of the radial velocities are used to refine the
background modelling which, in turn, is used to refine the estimates
of the radial velocities.

The simulations performed to compute Fig.~\ref{fig:crowding} assume
that the radial velocities are known with precisions of: $\sigma V_R =
2$ km~s$^{-1}$ at $V=14$ and 20 km~s$^{-1}$ at $V$~$=$~17 (this
simulates an intermediate stage of the iterative process).  The result
of the background modelling (grey lines in the top panels of
Fig.~\ref{fig:crowding}) is encouraging. First tests based on the
background modelling technique \citep{zwitter2003} inspire confidence
that radial velocities can be obtained from such overlapping spectra
with moderate degradation of performance up to $\simeq 20\
000$~stars/deg$^2$ with $V$~$<$~17.

The information obtainable from crowded regions will be worse than for
sparsely populated ones. Nevertheless, the overlapping stars degrade
the RVS accuracy mainly by increasing the background shot noise
level. Such a conclusion is justified if one can rely on the accurate
knowledge of the flux throughput of the instrument and on accurate
spectral modelling of the background stars. In forthcoming studies, we
intend to investigate this issue in detail and, in particular, to
implement and test the full iterative analysis process.

\subsection{Atmospheric parameters and individual abundances} \label{atmospheric}
{\it Gaia} will rely on a large variety of observables to classify and
parameterise the stars \citep{bailerjones2002,bailerjones2003}:
distances (and therefore absolute magnitudes), 5 broad photometric
bands, 11 medium photometric bands and $R=11500$ spectra over the
wavelength range [848, 874]~nm. In this section we focus on the
information contained in the spectroscopic data.

The spectral richness of the RVS wavelength range gives a broad
diversity of line absorption responses which vary with $\rm T_{eff}$,
$\log g$ and $[Z/Z_\odot]$\footnote{The variation of line intensity
responses with the atmospheric parameters are well detailed in Table~4
of \citet{cayrel1963}.}. As a consequence, the RVS spectral range
contains all the required information to derive the stellar
atmospheric parameters.

At present, the spectroscopic determination of the atmospheric
parameters requires the micro-turbulence velocity to be
constrained. In the future, this {\it ad hoc} parameter will be
replaced by a 3-dimensional description of the stellar atmospheric
structure (temperature, pressure {\it etc.}) and velocity field.

Many methods have been developed to determine stellar atmospheric
parameters from spectra. One widely-used technique is detailed stellar
spectral analysis. The effective temperature and surface gravity are
constrained by requiring that abundances derived from different lines
be independent of the line excitation potential (for a given element)
and of the element ionisation stage. This method (even fully
automated) is not suited for the analysis of RVS spectra because of
the absence of sets of weak unblended lines from given elements in two
different ionisation stages, which would allow the ionisation
equilibrium to be constrained. Alternative techniques have been
developed based on automatic comparison using the minimum distance
method between measured and synthetic spectral quantities such as the
equivalent widths (EW) of strong lines \citep{thevenin1983}.  These
methods also require a ``good'' S/N ratio, which reduces their
performance for faint stars.

For this reason, different techniques have been developed in order to
analyse spectra of moderate to low S/N ratios. In these techniques,
the set of EW values have been replaced by the whole spectrum
\citep{cayrel1991, thevenin1992, katz1998, allendeprieto2003},
diminishing the effect of a low S/N ratio and producing results
comparable to detailed analysis based on spectra with high
resolution. \citet{thevenin2003} have used a grid of synthetic spectra
to explore the influence of resolution on the recovered atmospheric
parameters of late-type stars. They concluded that resolutions
$R=$~12000 and higher were appropriate to recover the three main
atmospheric parameters. \citet{soubiran2001} has evaluated the
performance of the comparison of full spectra using the minimum
distance method in the RVS wavelength range and assuming a resolving
power half that adopted for {\it Gaia}. She degraded the observed
spectra from the library of \citet{cenarro2001} to a S/N=20 and
compared each spectrum to the rest of the library one-by-one. The
parameters of the stars were recovered with precisions of:
$\sigma(T_{eff})=275$~K, $\sigma(\log g)=0.36$,
$\sigma([Fe/H])=0.22$~dex. A S/N=20 by the end of the mission
corresponds to $V \simeq$ 14.5 for a K1III star, $V \simeq 14.0$ for a
G5V star and $V \simeq 13.5$ for an F2 II star. Because of its higher
resolution, the RVS performances are expected to be better than this.

Another way to extract the atmospheric parameters is to use Neural
Networks \citep{bailerjones2000}. Tests have demonstrated
\citep{bailerjones2003} that using the R=5800 library\footnote{The
median signal to noise ratio per resolution element of the library is
85.}  of \citet{cenarro2001}, $\rm T_{eff}$ can be determined to
within 5\%, $\log g$ within 0.5 (solar metallicity) to 1.0 (metal poor
stars) and [Fe/H] within 0.3 dex.

Up to $V \simeq14-15$, the atmospheric parameters will be derived
using not only the spectroscopic observations, but also (and jointly)
the astrometric and photometric data. The global precision of the
atmospheric parameters from {\it Gaia} should be higher than that
estimated from the spectroscopic information alone, which is presented
here. The astrometric and photometric observations will also allow the
evaluation of the atmospheric parameters of the fainter stars.

For the brightest stars, measurements of equivalent widths and/or
synthetic profile fitting will allow the extraction of individual
abundances from RVS spectra.  The elements concerned are mainly Fe,
Ca, Mg and Si for the late-type stars F, G, and K. The abundances of
other elements like N can be derived in hotter stars such as A-type
stars. For the cooler stars, K and M types, molecular bands can
provide information on C, N or TiO abundances. This raises the
possibility of investigations into the chemical evolution of the
Galaxy.  Due to the quality of the spectra, the error on the deduced
stellar abundance of a chemical element depends on the resolution, the
S/N and the spectral sampling on the CCD. Using Cayrel's formula
\citep{cayrel1988}, the error on an equivalent width measurement of
0.005~nm at a resolution $R = 11\ 500$ and for $S/N=$ 50, is
0.0015~nm, corresponding, for this weak line, to an error of 0.13 dex
on the abundance.  A S/N=50 (at the end of the mission) corresponds to
$V \simeq$~13 for G5V and K1III stars and to $V \simeq$ 12 for an F2
II star.

The RVS spectra will be used, jointly with the photometric and
astrometric data, to determine the atmospheric parameters of about
$10-25\times10^6$ stars down to $V \simeq 14-15$. They will also
permit the extraction of individual abundances in $2-5\times10^6$
stars down to $V \simeq 12-13$.

\begin{figure*}
\centering
\resizebox{\hsize}{!}{
\includegraphics[width=8cm]{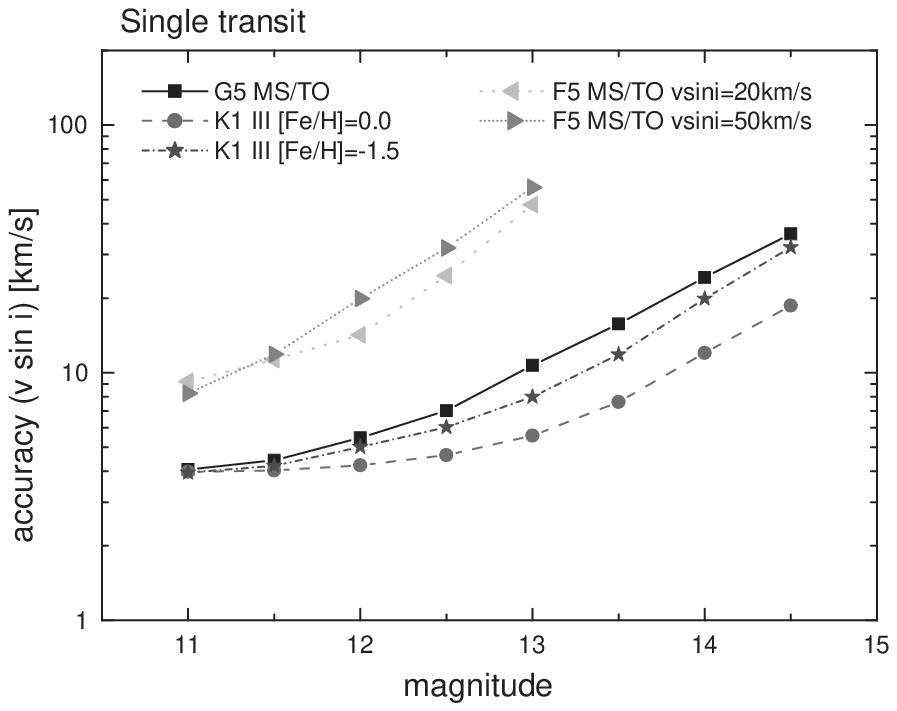}
\includegraphics[width=8cm]{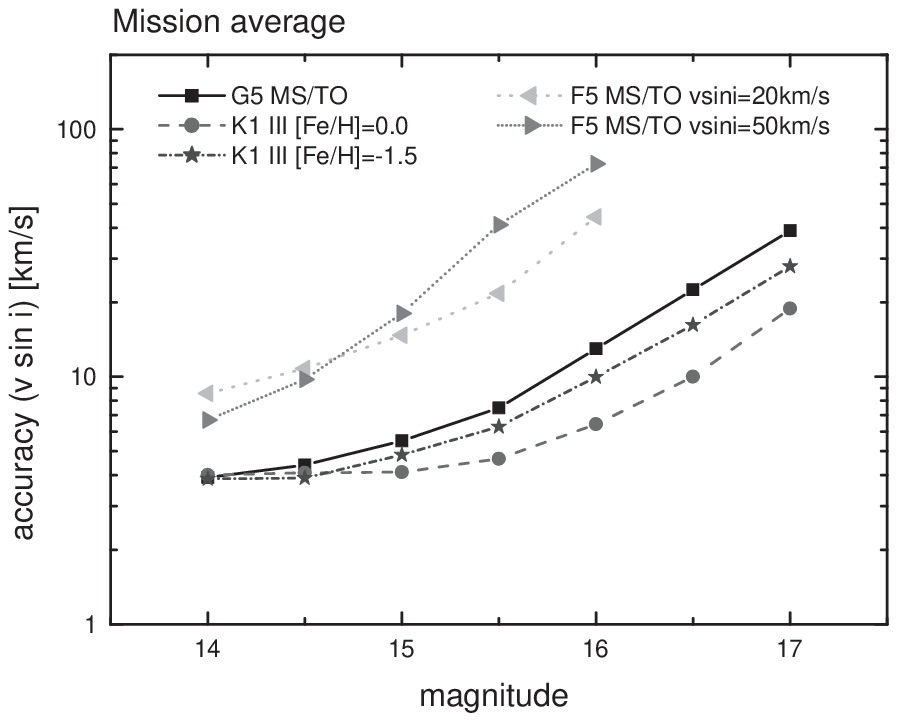}}
\vskip -1.2 cm
\resizebox{\hsize}{!}{
\includegraphics[width=8cm]{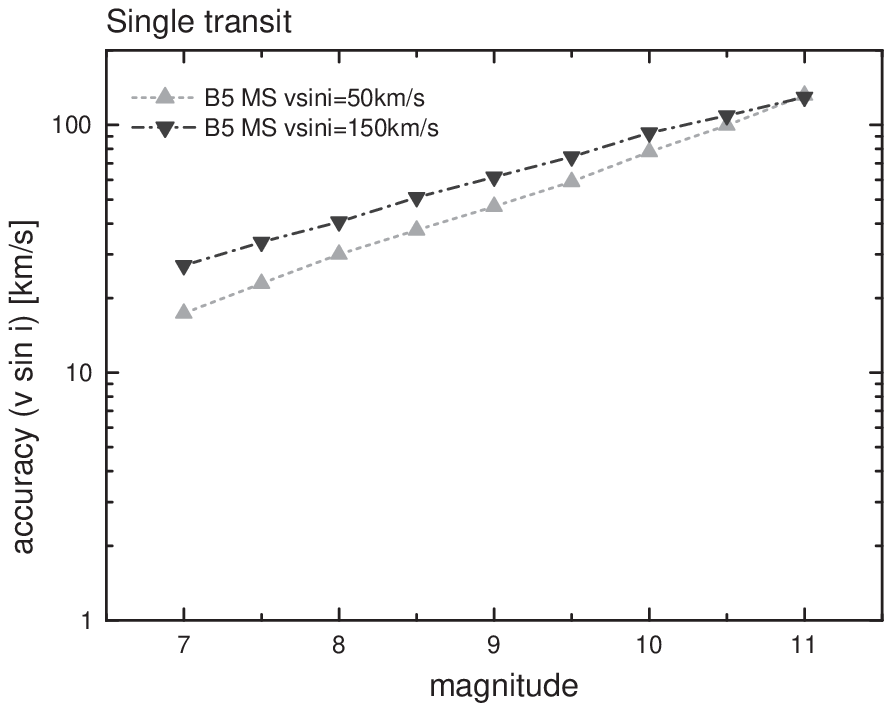}
\includegraphics[width=8cm]{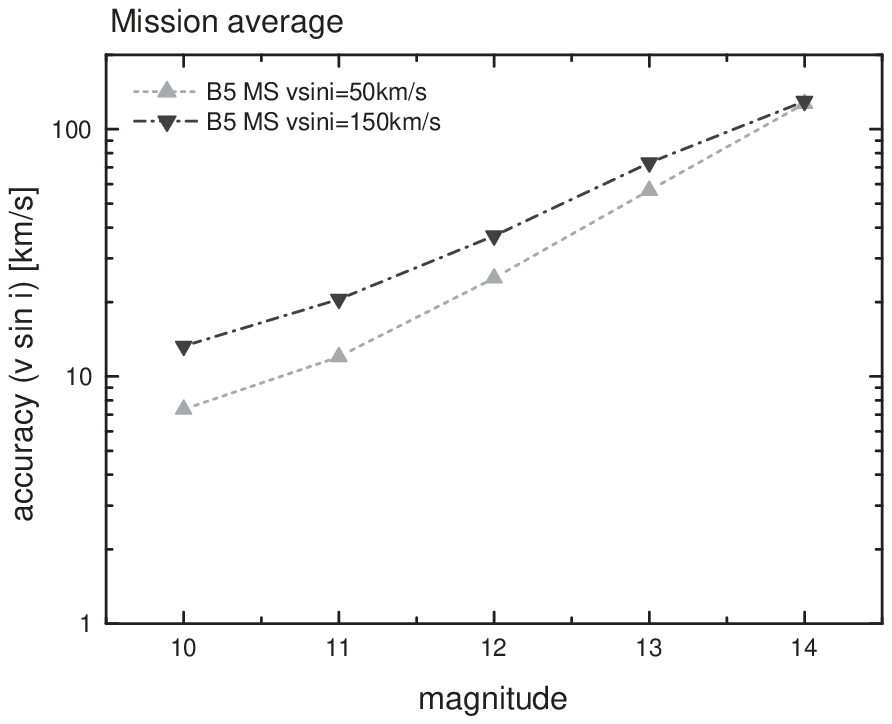}}
\caption{Precision of $v \sin i$ for a single stellar transit over the
RVS focal plane (left) and at the end of the mission (right) as a
function of magnitude and for the stellar types: K1 III, G5 MS/TO and
F5 MS/TO (top) and B5 MS (bottom).}
\label{fig:vsini}
\end{figure*}

\subsection{Rotational velocities} \label{rotational}
The analysis of stellar spectra obtained with {\it Gaia}-RVS will, in
addition to other stellar characteristics, enable the determination of
the projected rotational velocity ($v \sin i$) of individual stars.
The impact of the {\it Gaia} contribution to stellar rotation
astrophysics depends primarily on the precision obtained on $v \sin i$
and on the dependence of this precision on stellar type and magnitude.

The spectroscopic determination of $v \sin i$ is based on the
rotational broadening of spectral lines. Analysis methods used to
determine $v \sin i$ include fitting by least squares, measurement of
FWHM, estimations of the width of the correlation peak, deconvolution
and Fourier transforms.

In order to estimate the precision of rotational velocities obtained
from {\it Gaia} spectra, we performed simulations on synthetic Kurucz
stellar spectra and used the simple least squares fit method.  The
simulations and results presented here are for the RVS spectral
resolution R=11500 (for results obtained for R=5000 - 20000 see
\citet{gomboc2003}).

The simulation starts by choosing a stellar type ({\it i.e.} a
spectrum with given $\rm T_{eff}$, $\log g$, $[Fe/H]$, $[\alpha/Fe]$)
and original $v \sin i$. This spectrum is used to generate RVS-like
spectra using the RVS simulator described in \S~\ref{radial}.
Afterwards, comparisons between simulated and noise-free templates
with various $v\sin i$ are performed and the template with the minimum
square deviation is taken as the best fit. The precision is taken to
be the standard deviation of the differences between the original and
recovered $v \sin i$ (number of trials N=1000).

To estimate the {\it Gaia}-RVS capabilities, simulations were
performed for 5 out of the 6 stellar types representative of the
different Galactic populations listed in Table~\ref{tab:tracers}. The
F2 supergiant was not considered in this study, because the
macro-turbulent motion in its pulsating atmosphere affects the line
profiles, which cannot therefore be used to estimate the rotational
velocity.

As the first step towards a detailed and more accurate estimation of
the precision of $v \sin i$ determinations, only single parameter fits
were performed. This is the best-case scenario, because it is assumed
that the other stellar parameters ($\rm T_{eff}$, $\log g$, $[Fe/H]$,
$[\alpha/Fe]$, $V_{r}$) are exactly known and in the fitting procedure
spectra which differ only in $v \sin i$ are used. The results are
presented in Fig.~\ref{fig:vsini}. Precisions of $\sigma_{v \sin i}
\simeq$~5~km~s$^{-1}$ should be obtained at the end of the mission for
$V \simeq 15$ late-type stars. For B5 MS stars, precisions (at the end
of the mission) of $\sigma_{v \sin i} \leq$ 10-20 km~s$^{-1}$ should
be obtained up to $V \simeq$ 10-11.

The distributions of estimated $v \sin i$ exhibit a small bias of a
few kilometres per second at the ``bright'' end, which increases at
fainter magnitudes. The errors due to the bias are included in the
performances presented in Fig.~\ref{fig:vsini}.

In practice, the estimates of $\rm T_{eff}$, $\log g$, $[Fe/H]$,
$[\alpha/Fe]$ and $V_{r}$ obtained by {\it Gaia} will have their own
uncertainties, which will influence the precision of the $v \sin i$
determination. To estimate the influence of those uncertainties on the
precision of the estimated rotational velocity, similar simulations
were performed but the original spectrum was fitted with spectral
templates that had one of the parameters ($\rm T_{eff}$, $\log g$,
$[Fe/H]$, $[\alpha/Fe]$, or $V_{r}$) offset from its true
value. Results are still preliminary, but suggest that an offset of
0.5 in $\log g$ does not substantially influence the precision of the
estimated rotational velocity; the precision of $v \sin i$ is not
significantly affected by a mismatch of up to 125~K in effective
temperature ; errors of 0.1 dex in metallicity do not seem to affect
significantly the precision of $v \sin i$, while a 0.25 dex mismatch
leads to degraded rotational velocities ($\sigma >$10 km~s$^{-1}$);
and errors in radial velocity of 10 km~s$^{-1}$ lead to errors in
rotational velocities of more than 10 km~s$^{-1}$.

The results presented in Fig.~\ref{fig:vsini} are considered only as a
first order estimation of the performance of {\it Gaia} in the field
of stellar rotation determination. These results were obtained using
synthetic stellar spectra and, therefore, modelling uncertainties and
chemical composition peculiarities were not taken into
account. Furthermore, in regions of high stellar density, the effect
of crowding will also undoubtedly degrade the determination of the
rotational broadening of spectral lines.  The preliminary results also
show that the effects of the combined errors in template parameters
($\rm T_{eff}$, $\log g$, $[Fe/H]$, $[\alpha/Fe]$, $V_{r}$) are not
easy to predict. While all the results presented here were obtained by
single parameter ($v \sin i$) fitting, more accurate estimations of
the performance will be obtained in multi-parameter fitting
simulations.

The performance of other analysis techniques will also be assessed in
future work. For example, cross-correlation techniques would allow us
to disentangle the combined effect of rotational broadening, [Fe/H]
and radial velocity, which respectively influence the width, the area
and the position of the cross-correlation function.  Moreover, using a
box-shaped template would make the determination less sensitive to
spectral mismatch.

\subsection{Interstellar reddening} \label{interstellar}
{\it Gaia} will accurately place all the observed stars in a
3-dimensional representation of the Galaxy. The synergy between
spectroscopy and multi-band photometry will allow {\it Gaia} to
quantify the reddening law and the amount of extinction of
interstellar origin in any direction and to any distance through the
Galaxy. Continuity of the reddening along the line of sight will serve
to isolate {\em deviant} points during the iterative building of the
3D Galactic extinction map. Most of these deviant points will either
be due to the use of incorrect models for the intrinsic energy
distributions or to circumstellar absorption.

A direct probe, independent of the modelling of the intrinsic energy
distribution of the targets, would be valuable in calibrating the zero
point and scale of the 3-dimensional extinction map during its
iterative assembly, in distinguishing between the alternative causes
of deviation mentioned above and in providing some clues to the nature of
the intervening absorbing material.

\begin{figure}
\centering
\resizebox{\hsize}{!}{
\includegraphics{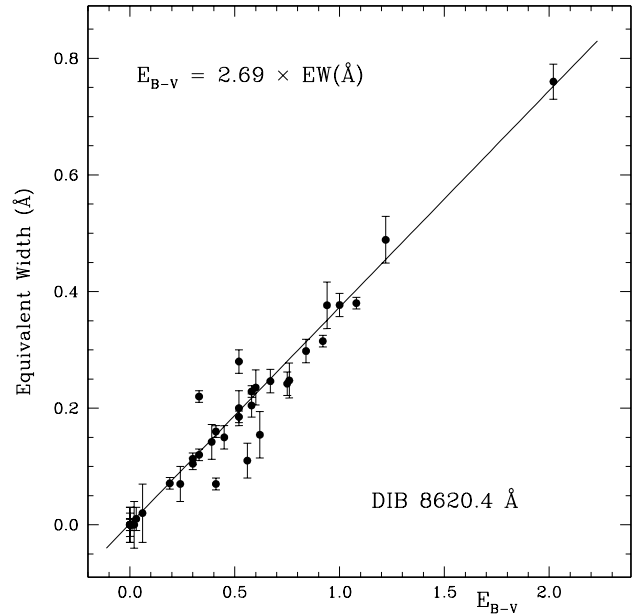}}
\caption{Correlation between the equivalent width of the 862.0~nm DIB
and reddening for stars covering a range of Galactic coordinates as
observed in Asiago (from Munari 2000).}
\label{fig:DIB}
\end{figure}

{\it Gaia} spectroscopy can provide such a direct probe. This is not
in the form of atomic absorption lines, however, because no
sufficiently strong resonant line lies within the {\it Gaia}
spectroscopic interval. Such atomic absorption lines would trace the
gaseous phase of the interstellar/circumstellar medium and a
gas-to-dust ratio would have to be adopted in order to convert it into
extinction. The interstellar medium, however, also manifests itself on
spectra with absorption features other than those of atomic
origin. These are termed {\em diffuse interstellar bands} (DIBs).

One such DIB lies right at the centre of the {\it Gaia}-RVS spectral
range, at 862.0~nm, where no significant spectral features appear in
the spectra of hot stars (cf. \citealt{munaritomasella1999}) -
the best suited to trace interstellar absorption to the
furthest distances. \citet{munari1999} was the first to investigate
this band in detail and found a remarkable correlation of its
intensity with reddening. Such a relation is unusual among DIBs, which
on average show only a mild correlation with reddening
\citep{herbig1995}.  \citet{munari2000} enlarged the sample of
investigated stars (covering a larger range of Galactic coordinates
and distances) and reinforced the evidence for such a relation, which
takes the form (Fig.~\ref{fig:DIB}):

\begin{equation}
E_{B-V} = 26.9 \times E.W.({\rm{nm}})                    
\end{equation}

were E.W. is the equivalent width of the 862.0~nm DIB. Such a tight
relation with reddening suggests a direct association between the
carrier(s) of the 862.0~nm DIB and the dust phase of the interstellar
medium. The dimensions and composition of interstellar dust grains
change through the Galaxy leading to the known effects of the 217.5~nm
bump or the ultraviolet slope of the extinction law. It may be expected
that these changes will affect the slope of Eq.(2) which can be
generalised to cylindrical coordinates mapping the 3D structure of the
Galaxy as:

\begin{equation}
E_{B-V} = \alpha(l,b,{\rm D}) \times E.W.({\rm{nm}})     
\end{equation}

Both the extinction and reddening law map the solid phase of the
interstellar medium ({\em i.e.} the dimensions and chemistry
of the dust grains). The nature of the DIB carriers is
still unknown, even if they are generally believed to be associated
with complex molecules, and this makes DIBs potentially useful to trace
(at least in part) the gaseous phase of the interstellar medium. It may be
expected that regions deviating from the dust characteristics of the
diffuse interstellar medium (as in some circumstellar nebulae and winds)
will display a 862\,nm DIB deviating from the above relation. In this
regard, while {\it Gaia} photometry will be by far the main tool for the
derivation of reddening information, the 862\,nm DIB appears to be a
useful probe of the conditions within the interstellar medium.
For example, HD~62542 has one of the most extreme ultraviolet extinction
curves known, featuring a very broad, shortwardly-displaced 217.5~nm bump
and an extremely steep far-UV rise.  In spite of an estimated $E_{B-V}$=0.35,
it does not show appreciable signs of classical 578.0, 579.7, 627.0, 628.4,
and 661.4~nm DIBs in its optical spectrum \citep{snow2002}.

An 862.0~nm DIB as weak as that corresponding to $E_{B-V}$=0.10 should
be easily identifiable on the $R$=11\,500 {\it Gaia}-RVS spectra of
all early type stars with S/N$\geq$100. This will provide a great
opportunity for the direct probing of the interstellar and
circumstellar medium, independent of the photometric approach.
Additional, dedicated, ground-based observations to investigate the
nature and behaviour of the 862\,nm DIB in more detail prior to the
{\it Gaia} launch are to be encouraged.

\subsection{Data archiving, calibration and analysis} \label{analysis}
The development of the on-ground data archiving and processing system
is one of the key challenges during the preparation for the RVS and
for {\it Gaia} in general. During the 5 years of the mission, {\it
Gaia} will collect about $10-15$ billion spectra of $\simeq 100-150$
million stars as well as astrometric data and multi-band photometry
for about a billion stars.  The calibration and analysis of such a
huge amount of heterogeneous data (all spectral types and luminosity
classes including variable and peculiar stars, single and multiple
systems, very different field stellar densities and therefore levels
of crowding, different temporal sampling as a function of Galactic
coordinates, three different instruments), as well as the particular
observing mode (continuous and repeated scans of the celestial sphere)
require new, robust and fully-automated methods to be developed.  The
RVS data will not be processed independently of the astrometric and
photometric observations: on the contrary, specific methods will be
developed to calibrate and analyse the {\it Gaia} data, as much as
possible, in a global way using the whole body of information provided
by the three instruments.

The ``{\it {\it Gaia} Database Access and Analysis Study}'' (GDAAS),
(2000-2004), is developing a prototype of the data archiving and
processing system. At the time of writing, mid-2004, a prototype of
the data model and archiving system is in place and several
algorithms, including the astrometric global iterative solution, have
been implemented and tested \citep{figueras2003}. By the end of the
study, more than 20 algorithms, including a non-optimised derivation
of the radial velocities, should be in place. The GDAAS activity will
be replaced from $\sim$2006 by the development of the full {\it Gaia}
archiving and analysis system.

In parallel with the development of data processing tools, studies are
in progress to improve our knowledge of atomic and molecular data, to
refine the computation of stellar models and synthetic spectra and to
build libraries of observed reference spectra. All this information
will be used by the calibration and analysis processes and their
quality will be directly reflected in the accuracy of the derived
stellar and interstellar parameters.

\section{Conclusions} \label{conclusion}
The Radial Velocity Spectrometer is a $2^\circ \times 1.61^\circ$
integral field spectrograph observing in TDI scan mode over the
wavelength range [848, 874]~nm with a resolving power $R = \lambda /
\Delta \lambda =11\ 500$. The RVS will provide a rich harvest of
stellar and interstellar parameters: radial velocities up to magnitude
$V \simeq$ 17-18, rotational velocities with precisions for late type
stars of $\sigma_{v \sin i} \simeq 5$ km~s$^{-1}$ at $V \simeq$~15,
individual element abundances (for stars with $V \leq$~12-13~mag) and
interstellar reddening.  It will also contribute, together with the
astrometric and photometric data, to the determination of the
atmospheric parameters of stars brighter than $V
\simeq$~14-15. Moreover, many stellar processes will imprint their
distinctive signatures in the RVS spectra: e.g. pulsation and
variability, mixing, accretion, winds and mass loss.  Finally, the
large number of epochs of observation will make it possible to detect
and characterise binary and multiple stellar systems as well as
periodic and transient phenomena.

During the two years 2001-2002, the RVS preparation studies were
mainly focused on the definition of the instrument scientific case and
on the comparison of the performance of the different possible
configurations. These studies converged in late 2002 with the
selection of the baseline concept. The RVS work is now mainly devoted
to (i) the optimisation and development of the instrument design, (ii)
the simulation of more realistic RVS-like spectra which will be used
to refine the evaluation of the performance of the spectrograph and
(iii) the definition and development of the on-board and on-ground
data processing algorithms.

\section*{Acknowledgments}
We would like to thank the ESA and industry teams preparing the {\it Gaia}
mission for their active and efficient support. We are very grateful
to R. Kurucz, F. Kupka, N. Piskunov and the VALD people for making their
software packages and molecular data available to the community.
DK, FT, FA, CT and SM acknowledge financial support from CNES.
MIW acknowledges financial support from PPARC. We are grateful to our
referee, G. Gilmore, for useful comments which improved the quality of
the manuscript.


\label{lastpage}

\end{document}